\documentclass[english,prl,aps,twocolumn,floatfix]{revtex4-1}
\usepackage{graphicx}
\usepackage{dcolumn}
\usepackage{amsmath,graphics,epsfig,epstopdf,color,verbatim,ulem}
\usepackage{amssymb}
\usepackage{babel}

\begin{document}

\title{Site-selective Mott transition in rare earth nickelates}

\author{Hyowon Park$^{1,2}$, Andrew J. Millis$^1$, Chris A. Marianetti$^2$}

\affiliation{
 $^{1}$Department of Physics, Columbia University, New York, NY 10027, USA \\
 $^{2}$Department of Applied Physics and Applied Mathematics, Columbia University, New York, NY 10027, USA
}

\date{\today}

\begin{abstract}
A combination of density functional and dynamical mean field theory calculations are used to show that the remarkable metal-insulator transition in the rare earth nickelate perovskites arise from a site-selective Mott phase, in which the $d$-electrons on a half of the Ni ions are localized to form a fluctuating moment while the $d$-electrons on other Ni ions form a singlet with holes on the surrounding oxygen ions. The calculation reproduces key features observed in the nickelate materials, including an insulating gap in the paramagnetic state, a strong variation of static magnetic moments among Ni sites and an absence of ``charge order''. A connection between structure and insulating behavior is documented. The site-selective Mott transition may be a more broadly applicable concept in the description of correlated materials.
\end{abstract}

\maketitle

Rare earth nickelates ReNiO$_3$ with Re=Sm, Eu, Y or Lu undergo a remarkable paramagnetic metal to paramagnetic insulator transition as the temperature is decreased below a temperature T$_{MI}\sim$ 400-600K~\cite{Alonso:01}. The insulating state is characterized by a two-sublattice symmetry breaking, with the Ni on one sublattice having a decreased Ni-O bond length and the Ni on  the other having an increased mean Ni-O bond length.   While the materials are sometimes classified as Mott/charge-transfer materials \cite{Torrance92,Imada98,Stewart:11}, the Mott picture does not account for the association of insulating behavior  with  bond disproportionation.  Charge ordering has been extensively discussed~\cite{Alonso99,Alonso:01,Staub:02,Mazin:07,Medarde:09}  but charge ordering in the naive sense of a change in the Ni valence between the two sublattices is expected to be suppressed by the large repulsive $d$-$d$ interaction $U$  on the Ni site and appears to be ruled out by very recent soft X-ray resonant diffraction data~\cite{Bodenthin:11}. Antiferromagnetic order can drive an insulating state and induce  a disproportionation as a second-order effect~\cite{Lee:11};  while this mechanism may be relevant to NdNiO$_3$, where the magnetic and metal-insulator transitions coincide, it fails to account for the paramagnetic insulating state  observed in  the Sm, Eu Y and Lu compounds.  The origin of the paramagnetic insulating state has heretofore remained mysterious. 

Here we present density functional plus dynamical mean field theory (DFT+DMFT) calculations which show that the bond-length disproportionation  and associated insulating behavior are signatures of a novel correlation effect in which the Ni $d$ electrons on one sublattice (Ni$_1$, long bonds)   become effectively decoupled from the surrounding lattice while the $d$ electrons on the other sublattice (Ni$_2$, short bonds) bind with holes on the oxygen sites to form a singlet state.  We term this state a {\em site-selective Mott insulator}. In the site-selective Mott state the singlet formation energy dictates the magnitude of the insulating gap, while the valence difference between the two Ni sites, of course nonzero by symmetry, is very small and is not relevant to the physics.

Our DFT+DMFT calculations are performed  using the Vienna ab-initio simulation package~\cite{Kresse19991758,Kresse199611169}  to obtain bands from which localized Ni $d$ orbitals (DMFT basis set) and O $p$ orbitals are constructed  using maximally localized Wannier functions~\cite{Wannier} defined over the full $\sim$10eV range  spanned by the $p$-$d$ band complex.  In most of our calculations we used the LuNiO$_3$ crystal structure with the $P2_1/n$ space group and lattice parameters  taken from a fully relaxed DFT+U calculation of the magnetically ordered state;  these are in close agreement with the experimentally measured parameters~\cite{Alonso:01} (See Table$\:$\ref{tab:relax})  for the antiferromagnetic and paramagnetic insulating states. In a few comparison calculations we used the  $Pbnm$ structure of the high-T phase of LuNiO$_3$. The LuNiO$_3$ crystal structure is such that  $e_g$ and $t_{2g}$  are no longer irreducible representations of the point group of either Ni site.  However, for each Ni the oxygen cage is sufficiently close to cubic that one can still choose a coordinate system  where there are a set of three and a set of two cubic harmonics which are closely grouped in energy,  and the off-diagonal elements are relatively small.   We determine the local coordinates for each $d$ Wannier function such that the sum of the square of off-diagonal terms in the $d$ manifolds on each site is minimized. 
We  refer to the approximately twofold (threefold) degenerate states as   $e_g$ ($t_{2g}$)  respectively.

We treat the filled $t_{2g}$ orbitals  with a static Hartree-Fock approximation (shown by recent work to be adequate except for fine details of the photoemission spectrum~\cite{Deng12})  while correlations in the Ni $e_g$ manifold are treated within single-site DMFT including the full rotationally invariant Slater-Kanamori interactions using $U$=5eV and $J$=1eV. The DMFT impurity problem is solved using the continuous time quantum Monte-Carlo method~\cite{Werner2006076405,Werner2006155107}. An important issue in the DFT+DMFT procedure is the `double counting correction'   which accounts for the part of $U$ already included in the underlying DFT calculation  and plays an important physical role by setting the mean energy difference between $d$ and $p$ bands   and thereby controlling the $d$ occupancy.    There is no exact procedure for determining the double counting correction;  we therefore consider a range of values, parametrized by the resulting $d$-occupancy and chosen  such that the $d$-valence within DFT+DMFT is not substantially different from the valence found  in the density functional calculations.  We find that changes to  $U$/$J$ or the double counting correction shift the phase boundary slightly but do not change the main results.  

\begin{figure}
\includegraphics[scale=0.45]{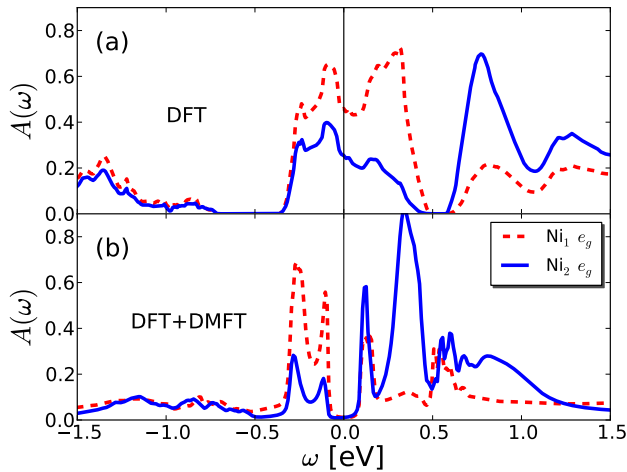}
\caption{
(Color online) Momentum-integrated spectral function $A(\omega)$ computed for $P2_1/n$ LuNiO$_3$, projected onto the local Wannier basis corresponding to the Ni $e_g$ states for Ni$_1$ (dashed, red) and Ni$_2$ (solid, blue) and displayed in the near fermi-surface frequency regime.
(a) DFT spectral function.
(b) DFT+DMFT spectral function obtained at $U$=5eV, $J$=1eV, $N_d$=8.0, and $T$=116K.
\label{fig:DOS}}
\end{figure}

Fig.$\:$\ref{fig:DOS} compares the $d$-electron spectral function for $P2_1/n$ LuNiO$_3$ obtained from a density functional calculation (Fig.$\:$\ref{fig:DOS}a) to that obtained from our DFT+DMFT calculation (Fig.$\:$\ref{fig:DOS}b). We see that while density functional theory predicts that the system is metallic (no gap at the chemical potential) despite the lattice distortion, the correlation effects captured by DMFT drive the system into an insulating state with a gap of $\sim$ 200meV (the precise gap value depends on the double counting prescription and the value chosen for $J$).
Differences in the near fermi-surface spectral functions for the two sites are evident in both cases, with the Ni$_1$ site having larger spectral weight in the low energy regime below the fermi level and the Ni$_2$ site having larger spectral weight above. This qualitative difference was interpreted by Ref.~\cite{Mazin:07} as evidence of charge ordering.  It is important to note, however, that the charge density is defined in terms of an integral of the spectral function over the entire frequency regime. We find that  the total $d$-charge (Wannier basis) is $8.22\pm0.08$ for the DFT spectra and $8.0\pm0.06$ for the DFT+DMFT spectra   in Fig.$\:$\ref{fig:DOS} (here the $+$ corresponds to Ni$_1$ and $-$ to Ni$_2$); our DFT+U calculations (atomic orbital basis, see Table$\:$\ref{tab:relax}) give an even smaller difference. 
Thus we conclude that the difference in charge between the two sublattices, while of course nonzero by symmetry, is not physically important and in particular is not the cause of the insulating behavior.

\begin{figure}
\includegraphics[scale=0.45]{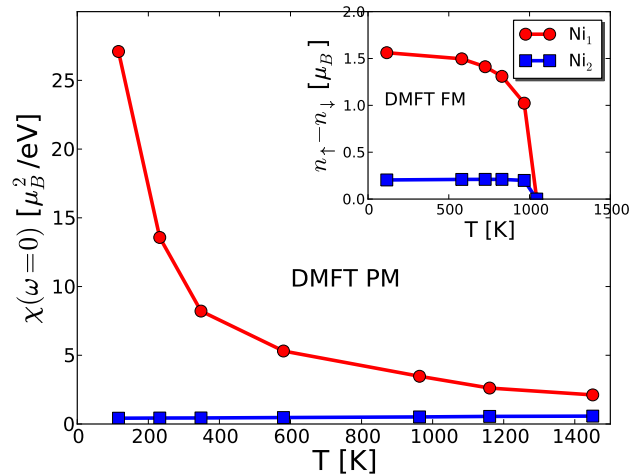}
\caption{
(Color online) Local magnetic susceptibility $\chi(\omega=0)$  of LuNiO$_3$ calculated via DFT+DMFT as a function of temperature in the paramagnetic state using the low $T$ $P2_1/n$ structure with mean bond length difference of 0.104$\AA$. Red circles, Ni$_1$ (large Ni-O bond length).  Blue square,  Ni$_2$ (small bond length). Inset: static magnetic moments $n_{\uparrow}-n_{\downarrow}$ from DFT+DMFT calculations. 
 \label{fig:Mom}}
\end{figure}

Fig.$\:$\ref{fig:Mom} shows the temperature dependence of the local magnetic susceptibility. A dramatic difference is visible between the two sites. The local susceptibility of the Ni$_1$ site has a clear Curie ($\sim 1/T$) behavior indicating well defined and long-lived but thermally fluctuating magnetic moments, as expected in a paramagnetic Mott insulator (in the single-site DMFT approximation). By contrast, the local susceptibility of the Ni$_2$ site has a negligible temperature dependence, indicating a lack of a local moment on this site. For this reason, we identify the origin of the insulating state as a {\em site-selective Mott transition}.

\begin{figure}
\includegraphics[scale=0.45]{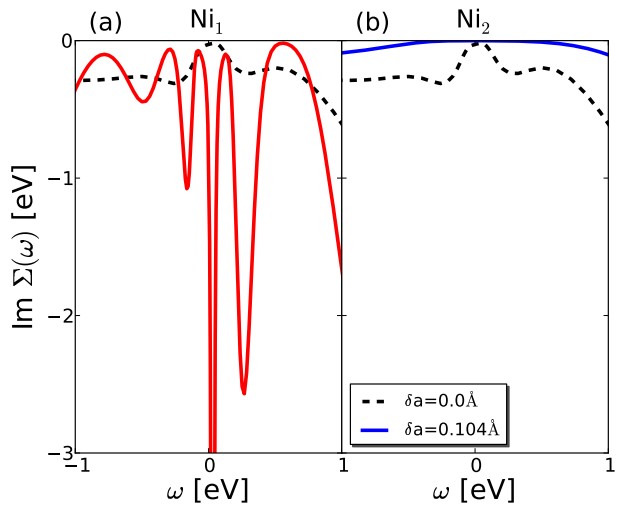}
\caption{
(Color online) The imaginary part of the self energies $Im\Sigma(\omega)$ of LuNiO$_3$ for the Ni$_1$ site (panel (a), red) and Ni$_2$ site (panel (b), blue) calculated via DFT+DMFT  in the paramagnetic state using the low $T$ $P2_1/n$ structure with mean bond length difference of 0.104$\AA$ and analytically continued following Ref.~\cite{Haule}. The self energies in the high $T$ $Pbnm$ structure with the mean bond length difference as 0.0$\AA$ (black dashed line) are also given for comparison.\label{fig:Sigma}}
\end{figure}

Further insight into the physics of the insulating state is obtained from the self energy, $\Sigma(\omega)$, shown in Fig.$\:$\ref{fig:Sigma}. In the high temperature $Pbnm$ structure the Ni$_1$ and Ni$_2$ atoms are equivalent. The imaginary part of  $\Sigma(\omega)$ (dashed lines, black on-line) shows the quadratic variation expected of a Fermi liquid while the real part (not shown here) has a linear frequency dependence at low energy indicating that the system is moderately correlated with a mass renormalization of 2.46,  consistent with optical data on paramagnetic metallic LaNiO$_3$~\cite{Ouellette:10}. When the size of the disproportionation is increased to a mean Ni-O bond length difference $\delta a= 0.104\AA$, comparable to the distortion in the experimental structure, the solution becomes insulating and the self energy changes. For the Ni$_1$ site, the imaginary part of the self energy displays a pole at zero energy, as expected for a Mott insulator. However, the self energy of the Ni$_2$ site displays a gap, with no pole, as expected in a singlet state with the quantum numbers of a trivial insulator.

The different behavior of the two sites can be understood by considering the  limit of extreme disproportionation, in which the Ni$_1$ site is completely decoupled from the surrounding oxygen ions while the Ni$_2$ site is strongly coupled. The mean Ni valence found in our and other DFT calculations is approximately $d^8$ (two electrons in the $e_g$ orbitals on each Ni and one hole per Ni  in the oxygen orbitals). On each Ni site,  the  Hunds coupling results in a high spin $d^8$ (total spin $S=1$) configuration. On the long-bond site a spin 1 local moment results; on the short-bond Ni$_2$  site the Ni spin-1 is strongly coupled to the two holes on the O sites of the octahedron, forming  a singlet, so the local susceptibility is $T$-independent, while  the corresponding self energy has a gap (as in Kondo insulators).  This model shows that an insulator can be obtained even though the variation in charge between the two Ni sites is negligible and   the oxygen hole density is  equally distributed over the oxygen network; hence this is not a charge ordering. A physical picture similar to what is discussed here was mentioned as one of several possibilities  in a Hartree-Fock study by Mizokawa, Khomskii and Sawatzky~\cite{Mizokawa00}, and has  been discussed by Freeland~\cite{Freeland}.

\begin{figure}
\includegraphics[scale=0.45]{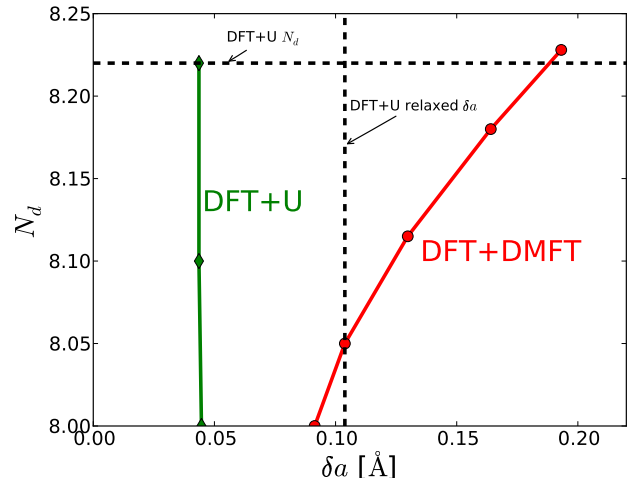}
\caption{
(Color online) The metal-insulator phase diagram of LuNiO$_3$ in the plane of bond disproportionation $\delta a$ and double counting (here parametrized by mean $d$-ocupancy $N_d$), obtained by DFT+DMFT (red circle) and DFT+U (green diamonds). The lattice disproportionation and mean $d$ occupancy  obtained from our  DFT+U calculation are shown as the vertical and horizontal  dashed lines respectively. 
\label{fig:Nd}}
\end{figure}

To determine the relation between the lattice distortion and the insulating phase, we studied a sequence of crystal structures interpolating between the high $T$ undistorted phase and the low-$T$ distorted structure, parametrizing the structure by the mean bond disproportionation $\delta a$.  Fig.$\:$\ref{fig:Nd} shows a phase diagram in the space of $\delta a$ and the double counting correction (here parameterized by the $d$-occupancy $N_d$) computed using  our DFT+DMFT procedure and, for comparison, 
via DFT+U which is (up to minor implementation differences not relevant here)  the Hartree-Fock approximation to DFT+DMFT.
We used the  VASP implementation of DFT+U,  but with with the double counting correction modified from the standard VASP  $E_{dc}^{\sigma}=U(N_d-0.5)-J(N_d^{\sigma}-0.5)$ ($\sigma$ is the spin index) to  
\begin{equation}
E_{dc}^{\sigma}=U(N_d-0.5)-J(N_d^{\sigma}-0.5)+E_{\Delta}
\end{equation}
In the VASP DFT+U implementation, the $d$-occupancy is fixed by a charge self-consistency loop; variation of $E_\Delta$ changes the final value, enabling computation of the phase diagram. 
The $d$-orbitals are defined slightly differently in the two methods (via maximally localized Wannier functions in DFT+DMFT and via projection onto atomic $d$ orbital functions defined within a sphere (radius set by the VASP default) around Ni atoms in DFT+U). 
As can be seen from Fig.$\:$\ref{fig:Nd} and by comparing the numbers in Table$\:$\ref{tab:relax} to those quoted above,  
these differences are not relevant to the physics we discuss here: the basic size of the $d$-occupancy, the difference
in $d$-occupancy between difference sites, and the variation with double counting correction are essentially the same. 


Fig.$\:$\ref{fig:Nd} shows that in the case of DMFT, changing the double counting correction so as to increase the $d$-valence away from $d^8$ increases the distortion magnitude required to obtain the insulating phase. The  DFT+U curve is much steeper, and favors  the insulating state, as expected from the Hartree-Fock nature of this approximation.

Finally, we consider the low temperature phase transition from the paramagnetic insulator to the magnetic insulator (AFI) (both having the $P2_1/n$ structure). The inset of Fig.$\:$\ref{fig:Mom} shows the DFT+DMFT results for the temperature dependence of the magnetic moments on the two sublattices as the temperature is decreased below the Curie temperature. The low $T$ limiting values are similar to experiment and to other calculations, but the magnetic transition temperature is grossly overestimated, presumably because of the neglect of spatial fluctuations and correlations in single-site DMFT~\cite{Jarrell}. Additionally, the ordered state is wrongly predicted to be ferromagnetic, as is also found in DFT+U calculations which will be analyzed further below.

\begin{center}
\begin{table}
\begin{tabular}{|c|c|c|c|c|}
\hline
   & Ni-O bond[$\AA$] & Mom[$\mu_B$] & $N_d$ & M/I \tabularnewline
\hline
   & 1.94/1.94 & & & \\
DFT & 1.96/2.01 & 0.74/0.73 & 8.21/8.20 & M \\
   & 2.01/1.96 & & & \\
\hline
   & 1.99/1.90 & & & \\
DFT+U & 2.02/1.94 & 1.52/0.57 & 8.22/8.24 & I \tabularnewline
   & 2.04/1.91 & & & \\
\hline
   & 1.92/1.92 & & & \\
DFT+U & 1.94/2.06 & 1.13/1.13 & 8.22/8.22 & M \tabularnewline
(Jahn-Teller)   & 2.06/1.94 & & & \\
\hline
     & 1.93/1.93 & & & \\
Exp. & 1.96/2.00 & N/A & N/A & M \\
($Pbnm$) & 2.00/1.96 & & & \\
\hline
     & 1.98/1.89 & & & \\
Exp. & 2.00/1.94 & N/A & N/A & I \\
($P2_1/n$) & 2.03/1.92 & & & \\
\hline
\end{tabular}
\caption{
The Ni-O bond length, magnetic moment, and $d$ occupancy for LuNiO$_3$ obtained from fully relaxed DFT and DFT+U structural
calculations. The  $N_d$ values are obtained by projecting the Kohn-Sham wavefunctions to atomic $d$ orbital functions defined within a sphere of radius 1.3$\AA$ around Ni atoms. For the DFT+U calculations, $U$=5eV and $J$=1eV are used. 
The Ni-O bond lengths for 
both the high-T ($Pbnm$) and low-T ($P2_1/n$) experimental structures~\cite{Alonso:01} are also given for comparision. 
\label{tab:relax} }
\end{table}
\end{center}

Although they cannot represent the paramagnetic insulating state, static mean field theories such as DFT, DFT+U and hybrid functional approaches may capture some of the physics of the antiferromagnetic insulating ground state.  Performing full structural relaxations within spin polarized DFT shows metallic behavior with neither magnetism nor   bond disproportionation (see Table$\:$\ref{tab:relax}): the relaxed DFT result closely matches the high temperature $Pbnm$ structure. This qualitative structural error  is uncommon for DFT and signals the importance of correlations.   We have also performed DFT+U calculations; this approximation stabilizes the disproportionated structure, giving structural parameters very similar to experiment.   The ground state of the DFT+U solution is found to be insulating and magnetic; the magnitude of the moments is substantially different between the two sites,  with the Ni atom with an expanded octahedron having a moment of $1.52\mu_B$ and the Ni atom with a contracted octahedron  having a moment of $0.57\mu_B$, similar to the DFT+DMFT calculation and to  what was found experimentally~\cite{Alonso99,Fernandez-Diaz01} and previously computed~\cite{Mazin:07}.  However, in both DFT+U and DFT+DMFT the magnetic ordering is found to be ferromagnetic; the experimentally observed antiferromagnetic state is metastable, being some 30meV/Ni higher in energy in DFT+U.  Furthermore, in the  DFT+U metastable antiferromagnetic solution   the smaller moment is strongly suppressed to $\sim 0.02\mu_B$, which is much smaller than the measured value.  Very recently, variational self-interaction-corrected density functional calculations~\cite{Puggioni:12} in LaNiO$_3$/LaAlO$_3$ superlattices were shown to produce the experimentally observed antiferromagnetic ground state within the nickelate layer, though the smaller moment is again much too small relative to experiment.  This suggests that non-local correlations may be necessary to properly describe the magnetic ordering, but we emphasize that the form of the ordered state is merely a low energy detail, clearly not responsible for the structural or metal-insulator transitions.  We also emphasize that in all of the methods and $d$ orbital definitions we have explored, the  $d$ charge difference between  the Ni$_1$/Ni$_2$ sites is found to be negligible: the insulating state should not be interpreted as charge-ordered.  Finally, within DFT+U there is a metastable state with an equivalent Jahn-Teller distortion on each Ni site  very close ($\sim$ 6meV/Ni) in energy. It is not impossible that this state could be realized under different circumstances.

In conclusion, we have identified a new mechanism for a correlation-driven metal-insulator transition, 
in which a translation symmetry breaking
lattice distortion leads to different spin physics (local moment vs singlet formation) on the
inequivalent transition metal sites of the insulator.
The mechanism is shown to  account for the essential properties of the paramagnetic insulating state of LuNiO$_3$ and related materials.  A key prediction of our work is a strong site dependence of the local magnetic susceptibility, with one sublattice exhibiting a Curie behavior and the other a temperature independent $\chi$.  The transition we have identified may be related to the insulating behavior observed in few-layer LaNiO$_3$-based heterostructures~\cite{Son:10,Chakhal:11} where DFT+U calculations~\cite{Chakhal:11} suggest that disproportionation can occur and drive a transition. 

$\textit{Acknowledgements:}$
The authors acknowledge funding from the U. S. Army Research Office via grant
No. W911NF0910345 56032PH.

\bibliography{nickelate}

\end{document}